\documentclass[letterpaper, 10 pt, conference]{ieeeconf}

\usepackage{pgfplots}
\pgfplotsset{compat=newest}
\usetikzlibrary{plotmarks}
\usetikzlibrary{arrows.meta}
\usepgfplotslibrary{patchplots}
\usepackage{grffile}
\usepackage{amsmath}
\usepackage{blindtext}
\usepackage{tabularx}
\usepackage[bottom]{footmisc}
\usepackage{cite}
\usepackage{url}
\newlength\fwidth
\usepackage[linesnumbered,ruled]{algorithm2e}

\usepackage{graphicx}
\usepackage{framed}
\usepackage{subcaption}
\usepackage{indentfirst}
\usepackage[]{algorithm2e}
\usepackage{multirow}

\usepackage{amsmath, amssymb,amsthm}
\usepackage{color}
\usepackage{booktabs}
\usepackage{amssymb}
\usepackage{mathtools,xparse}
\usepackage{physics}

\usepackage{comment}
\usepackage{hyperref}
\usepackage{cleveref}
\newtheorem{definition}{Definition}
\newtheorem{case}{Case}

\numberwithin{subcase}{case}
\newcommand{\transpose}{^{\mkern-1.5mu\mathsf{T}}}

\newcommand{\etal}{\textit{et al}. }
\newcommand{\ie}{\textit{i}.\textit{e}.}
\newcommand{\eg}{\textit{e}.\textit{g}.}
\renewcommand{\v}[1]{\mathbf{#1}}
\newcommand{\dtv}[1]{\dot{\mathbf{#1}}}
\newcommand{\LgP}{\v{L}_\v{g}\phi}
\newcommand{\LfP}{\mathrm{L}_\v{f}\phi}
\newcommand{\st}{\mathrm{s.t.}}
\newcommand{\gphi}{\nabla \phi\transpose (\v{x})}

\DeclareMathOperator*{\argmin}{arg\,min} 

\usepackage[font=small,labelfont=bf]{caption}

\crefname{figure}{Fig.}{Fig.}
\Crefname{figure}{Figure}{Figure}
\crefname{theorem}{Theorem}{Theorem}

\IEEEoverridecommandlockouts                              

\title{Safe Control Algorithms Using Energy Functions: \\ A Unified Framework, Benchmark, and New Directions}

\author{Tianhao Wei and Changliu Liu
\thanks{This is the extended version of a paper submitted to \textit{${58^{th}}$ Conference on Decision and Control} March, 2019; revised August, 2019.}
\thanks{This work was supported in part by the National Science Foundation under Grant \#1734019, and in part by Holomatic.}
\thanks{T. Wei and C. Liu are with the Robotics Institute, Carnegie Mellon University, Pittsburgh, PA, USA (e-mail: \tt\small twei2, cliu6@andrew.cmu.edu).}
}

\begin{document}

\maketitle

\begin{abstract}
    Safe autonomy is important in many application domains, especially for applications involving interactions with humans. Existing safe control algorithms are similar to one another in the sense that: they all provide control inputs to maintain a low value of an energy function that measures safety. In different methods, the energy function is called a potential function, a safety index, or a barrier function. 
    The connections and relative advantages among these methods remain unclear. This paper introduces a unified framework to derive safe control laws using energy functions. We demonstrate how to integrate existing controllers based on potential field method, safe set algorithm, barrier function method, and sliding mode algorithm into this unified framework. In addition to theoretical comparison, this paper also introduces a benchmark which implements and compares existing methods on a variety of problems with different system dynamics and interaction modes. Based on the comparison results, a new method, called the sublevel safe set algorithm, is derived under the unified framework by optimizing the hyperparameters. The proposed algorithm achieves the best performance in terms of safety and efficiency on the vast majority of benchmark tests. 
\end{abstract}

\section{Introduction}
Safe autonomy has become increasingly critical in many application domains. We should ensure not only the safety of the ego robot, but also the safety of other agents (humans or robots) that directly interact with the autonomy. For example, robots should be safe to human workers in human-robot collaborative assembly; autonomous vehicles should be safe to other road participants.
For complex autonomous systems with many degrees of freedom, safe operation depends on the correct functioning of all system components, \ie, accurate perception, optimal decision making, and safe control. This paper focuses on safe control which is the last defense to ensure the safety of a system.

A safe control law needs to guarantee that the unsafe region of the system's state space is not \textit{reachable}. Additionally, it requires forward \textit{invariance} of the safe region in the sense that once entered, the state of the system will stay in the safe part of the state space. To achieve forward invariance or set-invariant control \cite{blanchini1999set}, a scalar energy function is usually designed such that the control objective (\eg, safety) is with low energy. Then the desired control law should drive the energy function in the negative direction whenever the system state is outside of the desired set (\eg, the safe region). An energy function has many other variations, \eg, a potential function, a barrier function, or a safety index. Representative methods include potential field method (PFM) \cite{khatib1986real}, sliding mode algorithm (SMA) \cite{gracia2013reactive}, barrier function method (BFM) \cite{ames2014control}, and safe set algorithm (SSA) \cite{liu2014control}.
Though the aforementioned methods all have similar structures in the sense that they provide control inputs to decrease the value of the energy function, the connections and relative advantages among them remain unclear. This paper introduces a unified framework for safe controllers to demonstrate how to interpret different safe algorithms as variants of a common energy-function-based control.

\begin{figure}[t]
    \begin{center}
        \includegraphics[width=0.45\textwidth]{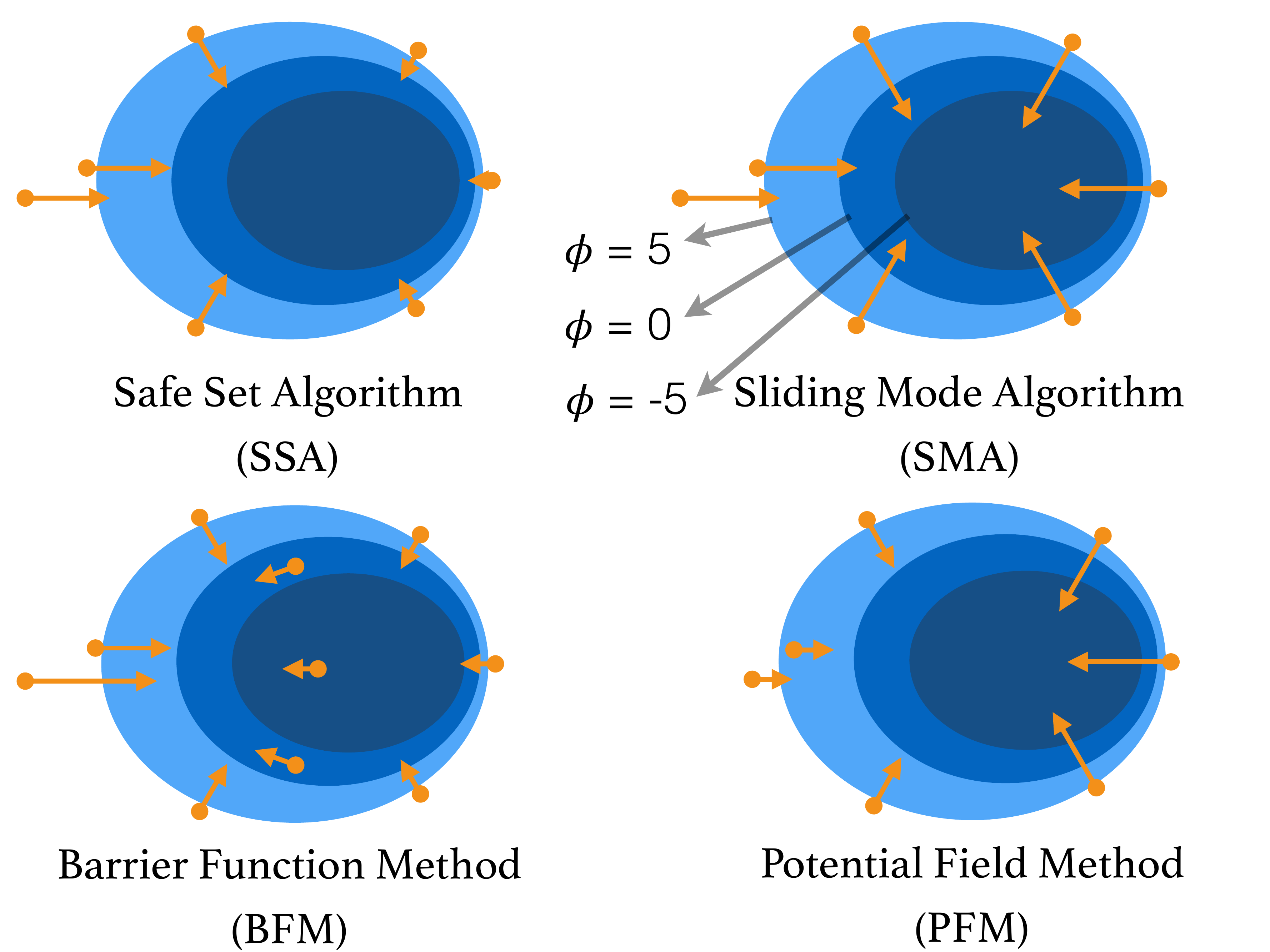}
        \caption{Illustration of different safe control algorithms on phase graphs. The plane represents the state space. The contours represent level sets of the energy function $\phi$. The system is safe when $\phi \leq 0$. An arrow indicates the direction and magnitude of the safe control input at a given state (dot).
1) SSA provides safe control inputs that are inversely proportional to the gradient $\nabla \phi$.
2) SMA provides safe control inputs along the gradient direction with constant magnitude.
3) BFM provides safe control inputs even when $\phi < 0$. The magnitude of control input depends both on $\phi$ and $\nabla \phi$. 
4) PFM provides safe controls that are proportional to $\nabla \phi$.
}
        \label{fig:safe_control}
        \vspace{-15pt}
    \end{center}
\end{figure}

Moreover, verification and validation of safe control algorithms are important. Lyapunov analysis \cite{branicky1998multiple} is usually adopted to prove invariance of the safe region using safe control laws. In addition to safety performance, we are also interested in understanding how conservative a control law is, and how much optimality or efficiency is sacrificed for the sake of safety. In general, it is difficult to mathematically analyze the trade-off between safety and efficiency in complex tasks. Empirical studies are needed to compare the performance of different algorithms in diverse complex situations. This paper introduces a benchmark system to test safe control algorithms on different dynamic systems (ball, unicycle, SCARA, 4 DoF robot arm) and different interaction modes (passive human model, interactive human model). We focus on three major metrics: safety, efficiency, and a hybrid score that incorporates both safety and efficiency. These metrics reflect the three most concerned aspects: human safety, robot efficiency, and robot safety. To the best knowledge of the authors, this is the first comprehensive benchmark on safe control. It can be used to evaluate not only energy-function-based safe control algorithms, but also controllers based on non-analytical methods such as those based on reinforcement learning~\cite{berkenkamp2017safe} and imitation learning~\cite{sun2018fast}. 

Based on theoretical analysis and comparison results of existing algorithms, it is observed that SSA and BFM have the best performance. Both SSA and BFM have relative advantages over the other in certain circumstances. SSA is triggered less frequently but reacts more radically, hence good for scenarios that are less safe. BFM reacts more gently but is triggered more often, hence good for scenarios that are safer. Based on the observation, we propose a new method, the sublevel safe set algorithm (SSS), combining the strengths of SSA and BFM. This method achieves the best performance on the vast majority of our benchmark tests.

The contributions of the paper are listed below.
\begin{enumerate}
\item This paper introduces a unified framework to derive safe control algorithms and shows that existing methods fit into the framework.
\item This paper develops a benchmark for existing safe control methods on a variety of problems with different system dynamic and different interaction modes.\footnote{The benchmark is available at \url{https://github.com/intelligent-control-lab/BIS}.}
\item This paper proposes a new safe control algorithm SSS that outperforms exiting algorithms on most benchmark tests.
\end{enumerate}

The remainder of the paper is organized as follows. \Cref{sec: problem} proposes a unified framework to derive safe control and shows how existing methods fit into the framework. \Cref{sec: benchmark} introduces the benchmark system. \Cref{sec: result} shows the comparison among different methods. \Cref{sec: discussion} discusses the results. \Cref{sec: conclusion} concludes the paper.

\section{Problem and Framework\label{sec: problem}} 

\subsection{Notation and Problem Definition}

Consider the following affine dynamical system with $n_x$ states and $n_u$ inputs
\begin{align}
    \dtv{x} = \v{f}(\v{x}) + \v{g}(\v{x}) \v{u} \label{eq: dynamics},
\end{align}
where $\v{x} \in X \subset \mathbb{R}^{n_x}$ is the state vector defined in configuration space, $\v{u} \in U \subset \mathbb{R}^{n_u}$ is the control input vector assumed to be unconstrained, $\mathbf { f } : \mathbb { R } ^ { n _ { x }} \rightarrow \mathbb { R } ^ { n _ { x } } $ and $\mathbf { g } : \mathbb { R } ^ { n _ { x } } \rightarrow \mathbb { R } ^ { n _ { x } \times n _ { u }} $ defines the system dynamics. 

The objective of the system can either be to track a trajectory or to regulate around a settle point. A reference control $\v{u}_0$ is provided to fulfill the system objective. In a safe environment, the robot can just execute $\v{u}_0$. Otherwise, the reference control $\v{u}_0$ may need to be modified by a safe control algorithm to prevent collisions with obstacles. The resulting safe control input $\v{u}$ depends on $\v{u}_0$.

The robot is occupying a certain region of the Cartesian space denoted as 
$C_r \subset \mathbb{R}^3$.
Similarly, the space occupied by the obstacle is denoted 
$C_o \subset \mathbb{R}^3$.
We denote $\v{c}_r$ as the closest point on the robot to the obstacle, $\v{c}_o$ as the closest point on the obstacle to the robot. Mathematically,
\begin{align}
    \v{c}_r, \v{c}_o = \argmin_{\v{c}_r^* \in C_r, \v{c}_o^* \in C_o} \norm{\v{c}_r^* - \v{c}_o^*}_2.
\end{align}

\begin{figure}[t]
    \begin{center}
        \includegraphics[width=0.4\textwidth]{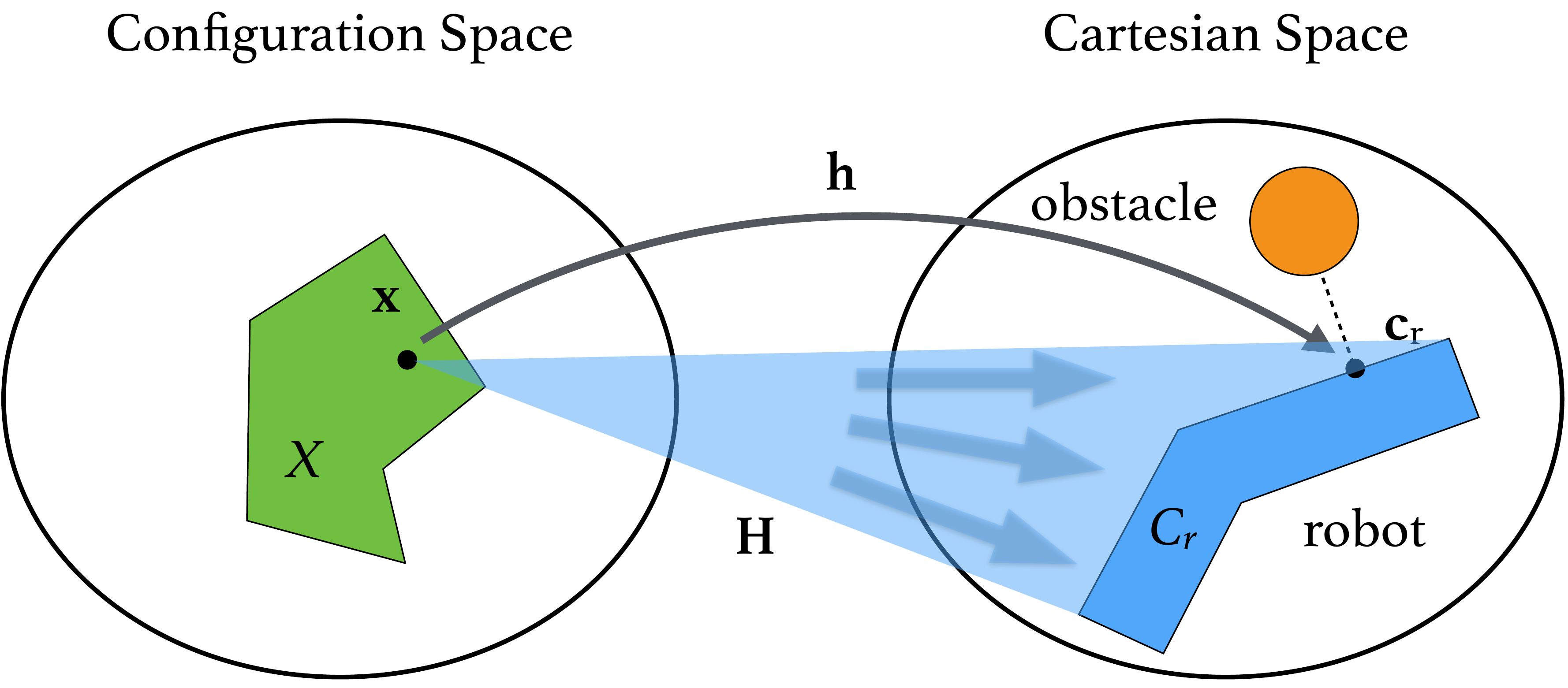}
        \caption{Illustration of the configuration space $X$, the state vector $\v{x}$, the occupied area $C_r$ in the Cartesian space, and the closest point $\v{c}_r$ on the robot to the obstacle.}
        \label{fig:X_to_Cr}
    \end{center}
    \vspace{-15pt}
\end{figure}

Let $\v{H} \colon \v{x} \mapsto C_r$ be a mapping from the robot state $\v{x}$ to its occupied region $C_r$.
Let $\v{h}  \colon \v{x} \mapsto \v{c}_r$ be a mapping from the robot state $\v{x}$ to the closest point $\v{c}_r$. 
The mappings depend on the robot model. The notations are shown in \cref{fig:X_to_Cr}.

The relation between time derivative of $\v{c}_r$ and time derivative of $\v{x}$ is
\begin{align}
    \dot{\v{c}_r} &= \v{h}'(\v{x})\ \dot{\v{x}} = \v{J}_{c_r}\ \dot{\v{x}}, \label{eq: c_h_x}
\end{align}
where $\v{J}_{c_r} = \v{h}'(\v{x})$ is the Jacobian matrix.

The relative distance between the obstacle and the robot is denoted by
$ d := \norm{\v{c}_r-\v{c}_o}_2$.
The relative velocity is denoted by
$ \dot{d} = \dv{}{t} \norm{\v{c}_r-\v{c}_o}_2$.
We define $d_{min}$ as the minimum required safe distance. 

\subsection{Energy Function and Control}

Energy-function-based methods use a customized energy function to measure safety. The lower, the safer. The value of $\phi$, which is usually called a safety index, increases when the robot is going toward the obstacle. The goal of the algorithm is to provide a control input that draws the robot away from the obstacle by decreasing the safety index.

We denote $\phi: X \rightarrow \mathbb{R} $ as an energy function defined on the configuration space and $\tilde{\phi}(\v{c}_r): C \rightarrow \mathbb{R}$ as an equivalent energy function on the Cartesian space, where
\begin{align}
    \phi(\v{x}) &= \tilde\phi(\v{c}_r) = \tilde\phi(\v{h}(\v{x})).
\end{align}
The function $\phi$ should be designed such that the system is safe if $\phi(\v{x}) \leq 0$. 
By the chain rule,
\begin{align}
    \nabla \phi(\v{x}) = \pdv{\phi(\v{x})}{\v{x}} = \v{J}_{\v{c}_r} \transpose\ \pdv{\tilde\phi(\v{c}_r)}{\v{c}_r} = \v{J}_{\v{c}_r} \transpose\ \nabla \tilde\phi(\v{c}_r). \label{eq: phi_x_c}
\end{align}

For safety, $\phi(\v{x})$ should be maintained negative. Once $\phi(\v{x})$ is high, \ie, in danger, it should be made decreasing, \ie, its time derivative $\dot\phi(\v{x})$ should be less than 0. Even in the safe situations, $\phi(\v{x})$ should not increase too fast, \ie, its time derivative may be upper bounded. Hence, we have an inequality constraint on $\dot\phi(\v{x})$ where 
\begin{subequations}\label{eq: dotphi}
\begin{align}
    \dot \phi(\v{x}) &= \gphi \ \dtv{x} \\
    &= \gphi \ (\v{f}(\v{x}) + \v{g}(\v{x})\ \v{u})\\
    &= \underbrace{\gphi \ \v{f}(\v{x})}_{\LfP}  + \underbrace{\gphi \ \v{g}(\v{x})}_{\LgP} \v{u} \leq \xi .\label{eq: constrain0}
\end{align}
\end{subequations}
The slack term $\xi\in\mathbb{R}$ is tunable. If $\xi > 0$, $\phi(\v{x})$ is allowed to increase within a certain rate. If $\xi<0$, $\phi(\v{x})$ must decrease. The inertia term $\LfP\in\mathbb{R}$ represents how the current state $\v{x}$ affects $\dot \phi(\v{x})$. The Lie derivative $\LgP\in\mathbb{R}^{n_u}$ represents how the control input $\v{u}$ affects $\dot \phi(\v{x})$.  

\subsection{Safe Control Algorithms}
Safe control methods that use energy-function-based approaches are reviewed below, in particular, the four methods shown in \cref{fig:safe_control}. These methods have different definitions of the energy functions. Since our framework focuses on the control strategies and can be generalized to any kind of energy function, we only review their control strategies below. Moreover, to focus on the main idea and reduce the number of hyperparameters in the unified framework, these algorithms are presented in their simplest forms. In \cref{sec: unified framework}, we show how these methods are related.

\subsubsection{PFM} Instead of deriving a control input $\v{u}$ in configuration space directly, PFM derives a control input $\v{u}_c$ in the Cartesian space first considering the following dynamics:
\begin{align}
    \dtv{c}_r = \v{u}_c := \v{u}_c^0 + \v{u}_c^*, \label{eq: PFM_1}
\end{align}    
where $\v{u}_c^0$ is the reference control in the Cartesian space transformed from $\v{u}_0$ based on \eqref{eq: dynamics} and \eqref{eq: c_h_x} such that
\begin{align}
    \v{u}_c^0 = \v{J}_{\v{c}_r}\ \v{g}\ \v{u}_0, \label{eq: PFM_2}
\end{align}
and $\v{u}_c^*$ is a repulsive ``force'' added to the reference $\v{u}_c^0$ whenever the safety constraint is violated. In particular, 
\begin{align}
    & \v{u}_c = \left\{\begin{array}{ll}\v{u}_c^0 - c_1\ \nabla \tilde\phi &\text{if } \tilde\phi \geq 0 \\
    \v{u}_c^0 & \mathrm{otherwise}
    \end{array}\right. \label{eq: pfc},
\end{align}
where $c_1>0$ is a tunable constant. 
Then the equivalent control input $\v{u}$ in the configuration space can be derived from $\v{u}_c$.

\subsubsection{SMA} It adds a correction term to the reference $\v{u}_0$ along the direction of the Lie derivative whenever the safety constraint is violated. 
\begin{align}
    & \v{u} = \left\{\begin{array}{ll}\v{u}_0 - c_2\ \LgP \transpose &   \text{if } \phi \geq 0 \\
    \v{u}_0 & \mathrm{otherwise}
    \end{array}\right. \label{eq: smc},
\end{align}
where the constant $c_2>0$ should be set large enough such that $\dot{\phi} = \LfP - c_2\ \|\LgP\|^2 + \LgP\ \v{u}_0$ is always negative. 

\subsubsection{SSA} It computes a control input that is closest to the reference $\v{u}_0$ and decreases $\phi$ when $\phi>0$.
\begin{align}\label{eq: SSA}
    & \v{u} = \min_\v{u} ||\v{u}_0 - \v{u}||_2,\ \st \ \dot \phi\leq \eta \ \text{or}\ \phi < 0,
\end{align}
where $\eta < 0$ corresponds to the slack term in \eqref{eq: dotphi}. 
SSA only deviates from the reference $\v{u}_0$ when $\phi \geq 0$. 

\subsubsection{BFM} It computes a control input that is closest to the reference $\v{u}_0$ and satisfies $\dtv{\phi} < \lambda\ \v{\phi}$ for a constant number $\lambda < 0$.
\begin{align}\label{eq: BFM}
    & \v{u} = \min_\v{u} \norm{\v{u}_0 - \v{u}}_2,\ \st\ \dtv{\phi} \leq \lambda\ \v{\phi},
\end{align}
where $ \lambda \v{\phi}$ corresponds to $\xi$ in \eqref{eq: dotphi}. 
BFM may always deviate from the reference $\v{u}_0$. 
When safe, \ie, $\phi < 0$, the control input may lead to the increase of the safety index $\phi$.

\begin{figure}[t]
    \begin{center}
        \includegraphics[width=0.4\textwidth]{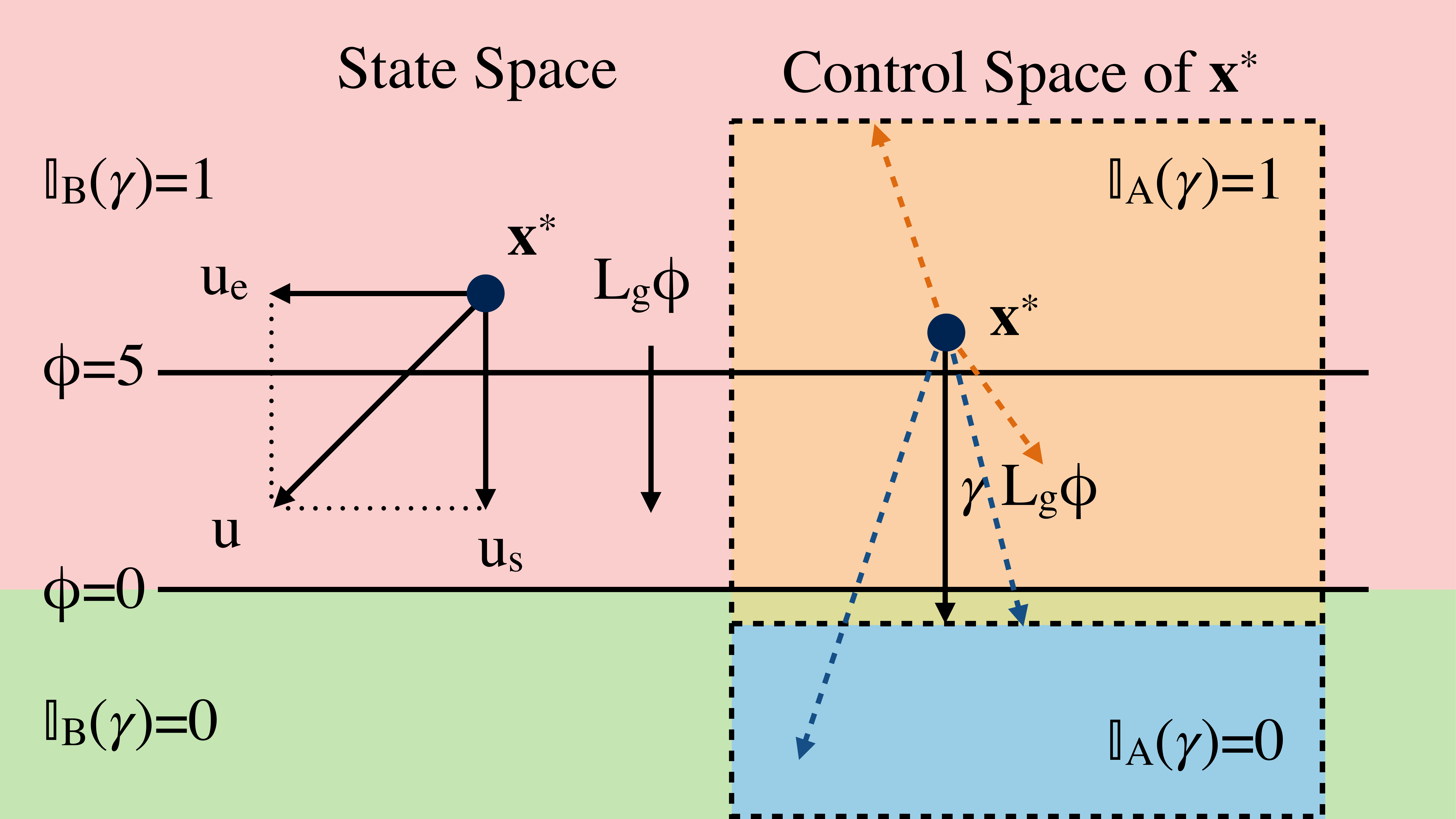}
        \caption{Illustration of the Lie derivative $\LgP$, perpendicular decomposition of $\v{u}$, and the indicator functions $\mathbb{I}_A(\gamma)$ and $\mathbb{I}_B(\phi)$. Left: Decomposition of $\v{u}$ along and perpendicular to $\LgP$. Right: Constraint on the control space. Blue arrows are examples of $\v{u}_0$ that comply with the constraint; Orange arrows are examples of $\v{u}_0$ that violate the constraint. It is assumed that $\dtv{x} = \v{u}$ in the figure.}
        \label{fig:decomposition}
    \end{center}
    \vspace{-15pt}
\end{figure}

\subsection{Unified Framework}\label{sec: unified framework}

Before proposing the unified framework, we first introduce a perpendicular decomposition of the control input $\v{u}$,
\begin{equation}
    \v{u} = \v{u}^s + \v{u}^e, \label{eq: u_decompose}
\end{equation}
where $\v{u}^s$ is parallel to $\LgP$, and $\v{u}^e$ is orthogonal to $\LgP$ as shown in \cref{fig:decomposition}. 
We call $\v{u}^s$ the \textit{safety component}, since the change of $\phi$ depends solely on $\v{u}^s$. We call $\v{u}^e$ the \textit{efficiency component}. When safety is ensured, we can add control input orthogonal to $\LgP$ to improve the efficiency of system performance.

Similarly, the reference control input $\v{u}_0$ can be decomposed as $\v{u}_0 = \v{u}_0^s + \v{u}_0^e$ where
\begin{align}
        \v{u}_0^s = \mu\ \LgP\transpose,
        \v{u}_0^e = \v{u}_0 - \v{u}_0^s,
\end{align}  
where 
\begin{equation}
\mu := \dfrac{\LgP\ \v{u}_0\ }{\norm{\LgP}^2}.
\end{equation}

For convenience, we introduce two indicator functions. Function $\mathbb{I}_B(\phi)$ indicates whether a state is dangerous, \ie,
    \begin{align}
        \mathbb{I}_B(\phi) &:= \begin{cases}
            1,\ \mathrm{if }\  \phi \geq 0 \\
            0,\ \mathrm{otherwise }.
        \end{cases} \label{eq: indicator_B}
    \end{align}

\
Function $\mathbb{I}_A(\gamma)$ is for optimization-based methods, \eg, SSA and BFM. It indicates whether $\v{u}_0$ violates the optimization constraint $\dot\phi \leq \xi$. According to \eqref{eq: dotphi}, the optimization constraint defines a half space in the control space as shown in \cref{fig:decomposition}, whose normal direction is along $\LgP$. Define 
        \begin{align}\label{eq: gamma defn}
            \gamma :=  \frac{\xi - \LfP}{\norm{\LgP}^2}.
        \end{align}
        Then an input $\v{u}$ satisfies the optimization constraint \eqref{eq: constrain0} if and only if $ \LgP\ \v{u} \leq \gamma \norm{\LgP}^2$. Hence, $\v{u}_0$ is feasible with respect to the constraint if and only if $ \LgP\ \v{u}_0 \leq \gamma \norm{\LgP}^2$ or equivalently $\mu \leq \gamma$. Then we define \footnote{There is a typo in the conference version,  ``\textgreater" should be ``\textless". \cref{fig:decomposition} is also remade because of this typo.}
    \begin{align}
        \mathbb{I}_A(\gamma) := \begin{cases}
            1,\ \mathrm{if }~ \mu < \gamma \\
            0,\ \mathrm{otherwise}.
        \end{cases}
    \end{align}

\begin{definition}[Energy-Function-Based Safe Control]
    A safe control method is called an energy-function-based safe control method if:
    \begin{enumerate}
        \item it uses a scalar energy function $\phi$ to measure safety;
        \item it provides safe control inputs in the following form 
        \begin{align}  \label{eq: safe control general form}
            &\begin{cases}
                \v{u}^s = \alpha\ \LgP\transpose \\
                \v{u}^e = \v{u}_0^e + \beta\ \v{u}_i^e
            \end{cases},
        \end{align}
        where $\alpha\in\mathbb{R}$ is a tunable parameter for safety response, $\beta\in\mathbb{R}$ is a tunable parameter for efficiency level, and $\v{u}_i^e\in\mathbb{R}^{n_u}$ is some vector orthogonal to $\LgP$.
    \end{enumerate}
\end{definition}

\newtheorem{theorem}{Theorem}
\begin{theorem}\label{main theorem}
    PFM in \eqref{eq: pfc}, SMA in \eqref{eq: smc}, SSA in \eqref{eq: SSA}, and BFM in \eqref{eq: BFM} are all energy-function-based safe control methods. They all satisfy \eqref{eq: safe control general form} with difference choices of parameters. In all methods, $\beta = 0$. 
    \begin{enumerate}
        \item for PFM: 
        \begin{equation}
            \alpha = \mu - \mathbb{I}_B(\phi)\ c_1 \label{eq: pf_a}.
        \end{equation}
        \item for SMA:
        \begin{equation}
            \alpha = \mu - \mathbb{I}_B(\phi)\ c_2 \label{eq: sm_a}.
\end{equation}
        \item for SSA:
        \begin{equation}
        \alpha = (1 - \mathbb{I}_A(\gamma)\ \mathbb{I}_B(\phi) )\ \mu + \mathbb{I}_A(\gamma)\ \mathbb{I}_B(\phi)\ \gamma \label{eq: ss_a},
        \end{equation}
        where $\gamma$ follows from \eqref{eq: gamma defn} and $\xi = \eta$.
        \item for BFM:
        \begin{align}
            \alpha = (1 - \mathbb{I}_A(\gamma))\ \mu + \mathbb{I}_A(\gamma)\ \gamma \label{eq: bf_a}
        \end{align}
where $\gamma$ follows from \eqref{eq: gamma defn} and $\xi = \lambda\phi$.
    \end{enumerate}    
\end{theorem}


\subsection{Sublevel Safe Set Algorithm}

Looking into the control strategies for different methods, we notice that 
\begin{enumerate}
    \item BFM's slack term in \eqref{eq: BFM} is a dynamic term that is related to energy function value, while SSA's slack term in \eqref{eq: SSA} is not.
    \item SSA may only provide control correction when $\phi>0$. The corresponding hyperparameter is \eqref{eq: indicator_B}. Yet BFM may provide control correction regardless of the value of $\phi$.
\end{enumerate}

When the parameters (\eg, the energy function $\phi$) are designed less conservative, SSA only starts to provide control correction at a close distance, while BFM may deviates from the optimal reference all the time. In this situation, SSA will be more efficient. When the parameters are designed more conservative, though BFM's control correction goes into effect from a far distance, BFM only makes a minor correction for most of the time. Meanwhile, SSA still provides radical corrections. In this way, BFM will be more efficient. These analyses will be supported by experimental results in \cref{sec: result}.

\begin{figure}[t]
    \centering
    \begin{minipage}[b]{.24\textwidth}
        \centering
        \begin{center}
            \includegraphics[width=\textwidth]{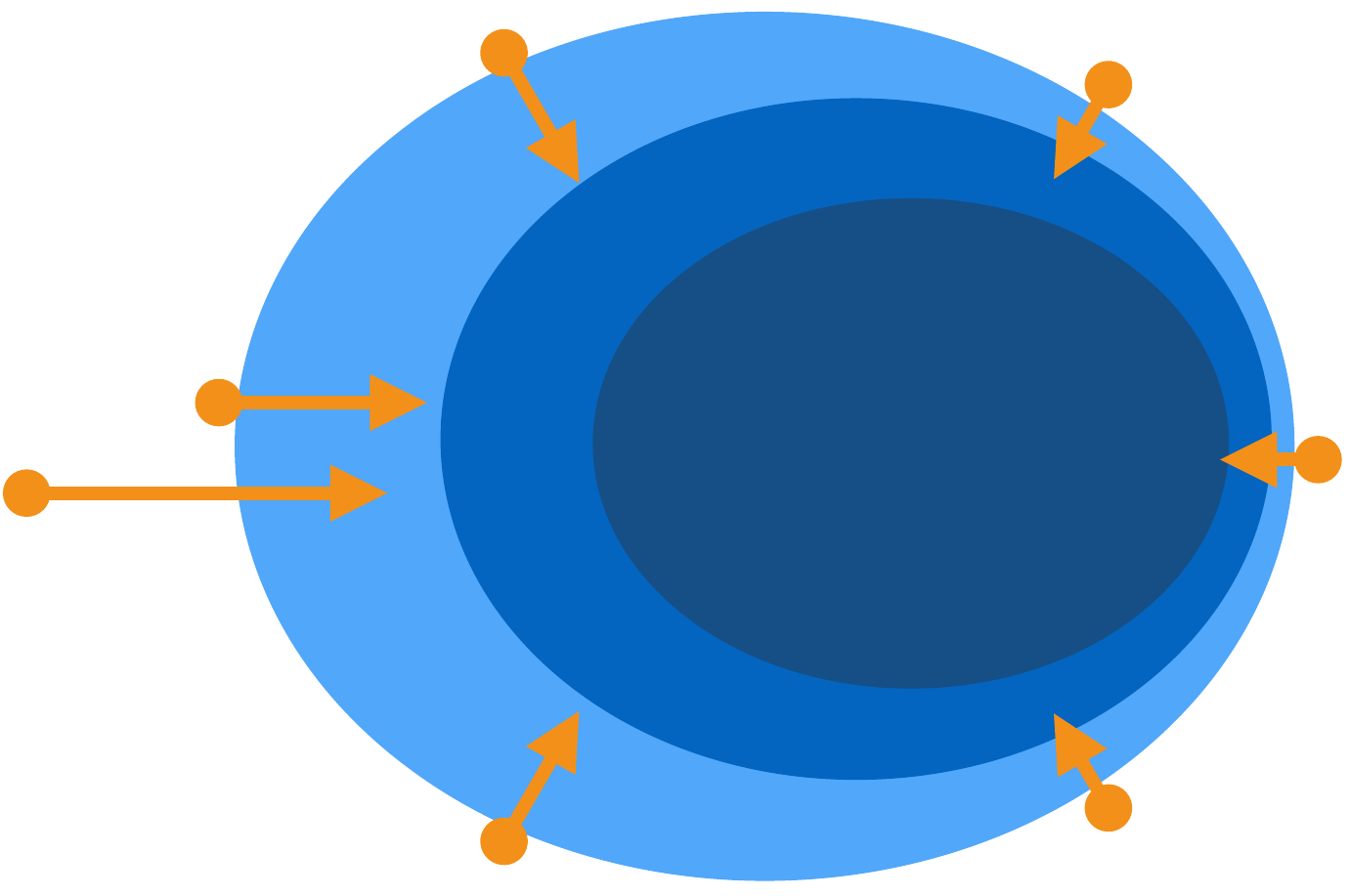}
            \caption{Phase graph for SSS.}
            \label{fig:SSS_phase}
        \end{center}
    \end{minipage}%
    \begin{minipage}[b]{.25\textwidth}
        \centering
        \begin{center}
            \includegraphics[width=\textwidth]{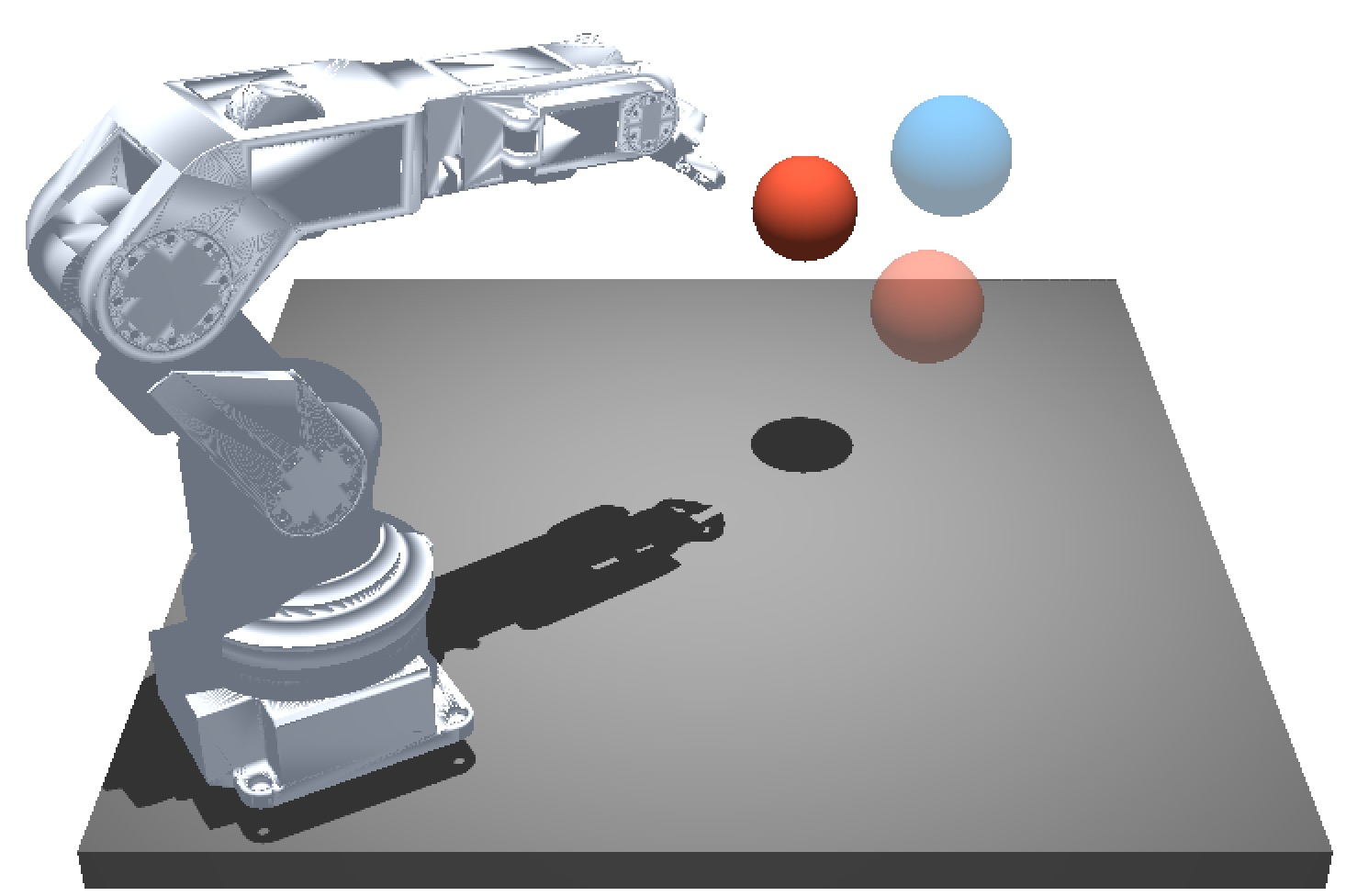}
            \caption{Screenshot of BIS.}
            \label{fig:BIS_screenshot}
        \end{center}
    \end{minipage}%
    \qquad
    \vspace{-15pt}
\end{figure}

We propose a new method, Sublevel Safe Set (SSS), to combine the strengths of SSA and BFM. SSA only provides control correction when $\phi\geq 0$, while the control correction relies on the value of the energy function $\phi$. The phase graph for SSS is shown in \cref{fig:SSS_phase}. Thus, no matter how the  parameters are designed, SSS should be more efficient than both SSA and BFM, which will also be verified in the experiment results in \cref{sec: result}.

By combining the slack term $\xi$ in BFM and the parameter $\alpha$ \eqref{eq: ss_a} in SSA, hyperparameters for SSS are designed to be
\begin{align}
    \begin{cases}
        \alpha = (1 - \mathbb{I}_A(\gamma)\ \mathbb{I}_B(\phi) )\ \mu + \mathbb{I}_A(\gamma)\ \mathbb{I}_B(\phi)\ \gamma \\
        \beta = 0\\
        \gamma = \frac{\xi - \LfP}{\norm{\LgP}^2}\\
        \xi = \lambda\phi
    \end{cases}.
\end{align}

The corresponding control strategy in its optimization form is
\begin{align}
    & \v{u} = \min_\v{u} ||\v{u}_0 - \v{u}||,\ \st \ \dot \phi< \lambda\ \phi \ \text{or}\ \phi < 0.
\end{align}

\section{Benchmark of Interactive Safety}\label{sec: benchmark}
\subsection{Overview of the Benchmark}
The unified framework provides a powerful tool to analyze connections and differences among algorithms. However, it is insufficient to solely rely on mathematical analysis to derive the performance of different algorithms in stochastic environments. In this paper, we are interested in the trade-off between safety and efficiency when these algorithms are applied on interactive tasks. Empirical studies are needed to compare different algorithms and understand their relative advantages in diverse complex situations. 

We introduce the benchmark of interactive safety (BIS) to perform empirical studies on safe control methods. BIS consists of a collection of robot models which can be used to benchmark and compare the performance of different safe control algorithms. A set of ready-made robot models can be easily extended by users. By changing robot models, we can test an algorithm with different system dynamics. By replacing the robot controller, we can test different control algorithms on a same dynamic system. By instantiating two robot models and letting a human subject (or a human-like controller) control one of them, we can test human-robot interactions. This paper is focused on the two agent case with one human and one robot, while multi-agent cases will be studied in the future.

In the simulation, the human agent is represented as an orange ball. The task of both the human and the robot is to reach a series of goals, which is represented by transparent balls. Screenshot of the scenario is shown in \cref{fig:BIS_screenshot}. 
BIS includes a data generator to generate test scenarios and an evaluator which evaluates all algorithms under the same condition.




\subsection{Robot Models under Comparison}
All robot models contains two modules: control module and execution module. In the control module, the robot first uses a Kalman filter to update the state of itself and the environment based on the measured data. Then the robot performs a collision check and computes the minimum distance and the closest point to the obstacle. Finally, the robot calls a control algorithm to compute the desired control input. 
In the execution module, the control input is applied to the robot simulator. The configuration of the robot is updated according to the dynamic model specified in the robot simulator. The robot simulator can also be replaced by a robot hardware to achieve hardware-in-the-loop evaluation.

BIS currently include four different robot dynamic models: ball robot model, unicycle robot model, SCARA robot model, and 4 DoF robot arm robot model. 


The robot model library in BIS can be easily extended. Adding a robot model into the robot model library requires two functions $\v{h}$ and $\v{h}'$. Function $\v{h}$ maps the robot state $\v{x}$ into Cartesian critical point $\v{c}_r$ as shown in \cref{fig:X_to_Cr}. Function $\v{h}'$ is the Jacobian in \eqref{eq: c_h_x}.

\subsection{Controllers under Comparison}
Controllers are called by robot models in the control module. BIS has included five energy-function-based control methods, \ie, PFM, SMA, BFM, SSA, SSS, and a human-like controller. The human-like controller models human behavior and generates control input that is similar to human. An imitation learning algorithm is designed to learn human behavior models from real human subjects. We asked 3 human subjects to control the human agent to achieve 100 goals one by one. The human model is learned from the demonstration data.

\subsection{Experiment Methods}

The experiments are setup with the following steps:
\begin{enumerate}
    \item Use a data generator to generate random goals and save them as test scenarios.
    \item Use the same test scenario and human model to test each algorithm.
    \item Compare the test results.
\end{enumerate}
We use 40 pieces of 30 seconds long test scenarios to test different algorithms in the experiments. The frame rate is 20 fps. 
In the study, it is assumed that 1) the Jacobian matrix does not change within one frame; and 2) all noises follow normal distribution.

The function $\phi$ is chosen to be \cite{liu2014control},
\begin{equation}
\phi = d_{min}^2 - d^2 - k \dot d.
\end{equation}

In the experiments, we test the performance of the algorithms under different values of their parameters. In particular, we tune the following two sets of parameters: 1) the parameters associated with $\phi$, \ie, $d_{min}$ and $k$; and 2) the parameters specific to each algorithms, \ie, $c_1$ for PFM, $c_2$ for SMA, $\eta$ for SSA, $\lambda$ for BFM, $\lambda$ for SSS.


\subsection{Evaluation Metrics}

Three metrics are used to evaluate the performance of different algorithms: an efficiency score and a safety score for human-robot interactions, and a hybrid score for robot co-working. For all scores, the higher, the better.

\subsubsection{Efficiency Score}
We use the average number of goals achieved in a given period by the robot as the efficiency score. 

\subsubsection{Safety Score}
Intuitively, this score is similar to the negation of a safety index. We design the safety score to be a weight sum of the relative velocity, where the weights are decided by the relative distance. The following factors need to be considered.
\begin{itemize}
    \item The score should decrease when the distance between the robot and the obstacle decreases.
    \item The score should decrease when the robot is moving faster to the obstacle.
    \item The weight should change rapidly when the robot is close to the obstacle.
    \item The score should not accumulate when the relative distance is larger than a threshold $d_s$.
\end{itemize}
Based on these considerations, the safety score is defined as
\begin{align*}
    \text{safety} = -\sum_0^{T} \min(0, \log(d / d_s))\ \dot d.
\end{align*}

The safety score takes both physical safety and psychological safety into consideration. When the robot is near the obstacle and rushing toward it, even if it does not end up with a collision, it receives a penalty because the robot is threatening the obstacle. 

\subsubsection{Hybrid Score}
In robot-robot collaboration scenario, physical safety is the only concern. Only collision matters, while psychological safety should be ignored. We define the hybrid score as the maximum efficiency without collision to evaluate the performance of the algorithms in these situations. This score reflects an algorithm's best performance when it can ensure physical safety.

\begin{center}
    \begin{figure*}[tb]
        \centering
        \includegraphics[width=\textwidth]{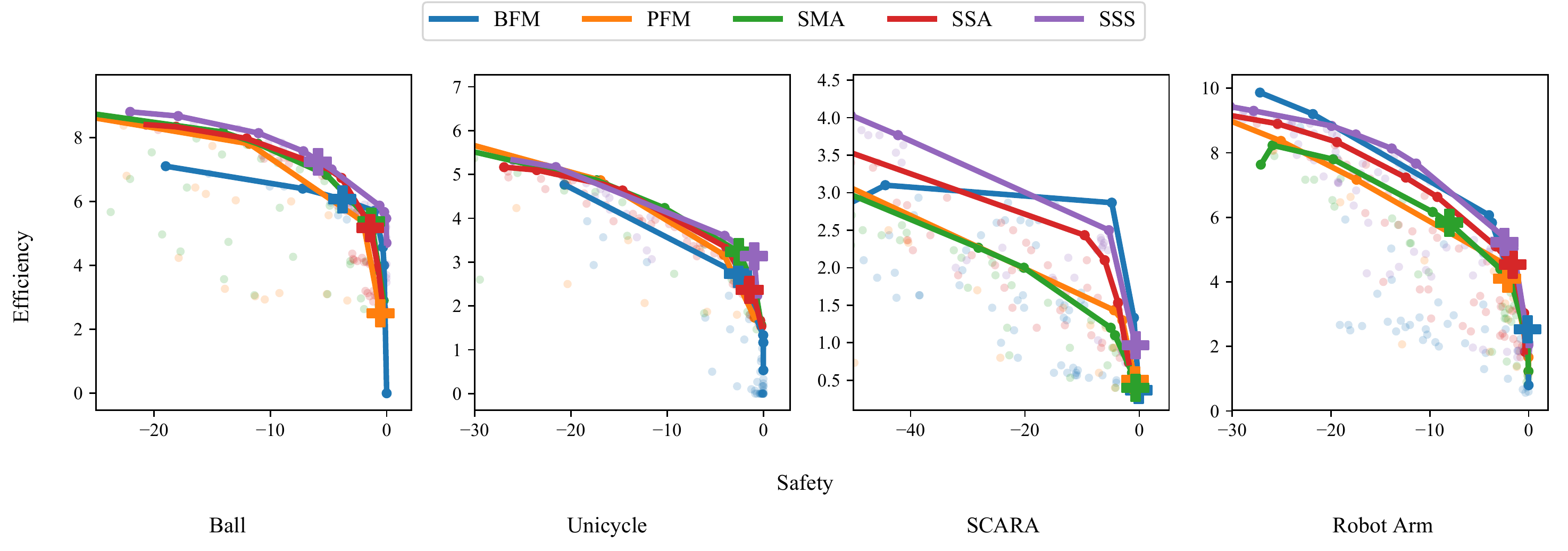}
        \caption[]
        {\small The trade-off curves between safety and efficiency for four robot models.} 
        \label{fig:trade off curves}
    \end{figure*}    
    \vspace{-15pt}
\end{center}

\section{Comparison Results\label{sec: result}}

\subsection{Trade-off between Safety and Efficiency}
We first evaluate the trade-off between the efficiency score and the safety score for all algorithms as shown in \cref{fig:trade off curves}. The trade-off curve can be obtained by tuning the parameters in the algorithms. For example, in PFM in \eqref{eq: pfc}, when the magnitude of $c_1$ is smaller, the system can be less safe but more efficient due to fewer detours. Different parameters may result in different safety and efficiency scores, which corresponds to different points on the trade-off graph in \cref{fig:trade off curves}. If a set of parameters leads to a high safety score, we say the parameters are conservative. The up-right convex hull of the those points is the trade-curve for one algorithm. If an algorithm has a trade-off curve covers all other algorithms, \ie, higher efficiency for the same safety score, we say it outperforms other algorithms.

SSA and SMA both provide control correction only in the boundary of the safe region of states. However, based on \eqref{eq: sm_a}and \eqref{eq: ss_a}, SSA provides a smoother control input comparing to SMA, which makes it more efficient. 



BFM provides control corrections even inside the safe region. This behavior makes it less efficient when the parameters are conservative. However, we noticed that BFM has a better performance than most algorithms with conservative parameters for the following two reasons.
\begin{enumerate}
\item When the parameters are conservative, control correction is triggered a lot. Frequent correction eliminates the advantage of only correcting at the boundary of the safe region, \ie, SSA and SMA are not superior at efficiency in this situation.
\item BFM's control correction is a dynamic term that is related to energy function value, while SSA is not. Though control correction is triggered more often, BFM only makes a minor correction for most of the time. However, SSA treats all corrections equally. In other words, SSA is more likely to be overreacting. 
\end{enumerate}

Our new algorithm SSS overcomes the drawbacks of SSA and BFM. Thus, it achieves the best performance on the vast majority of benchmark tests.

\subsection{Hybrid Score}

To demonstrate the performance of algorithms in a robot-robot collaboration scenario, we record the hybrid scores on different robot models as shown in \cref{table:hybrid score}. SSS has the best average performance, which achieves two best scores and two second scores. SMA achieves two best scores, which is out of our expectation.

\begin{table}[h]
\centering
    \begin{tabular}{lcccc}
    \toprule
            & Ball          & Unicycle      & SCARA         & RobotArm      \\
    \midrule
    SSS     & \textbf{7.23} & \underline{3.13}     & \textbf{0.96} & \underline{5.23}     \\
    BFM     & \underline{6.07}     & 2.73          & 0.37          & 2.53          \\
    SSA     & 5.17          & 2.37          & null     & 4.53          \\
    SMA     & 5.37          & \textbf{3.23} & \underline{0.39}          & \textbf{5.83} \\
    PFM     & 2.50          & null          & 0.03          & 4.10\\
    \bottomrule
    \end{tabular}
    
    \caption{Comparison of hybrid scores (Maximum Efficiency without Collision). The best two results are shown in \textbf{bold} and \underline{underline} respectively. If collision happens, the method gets a null as the hybrid score.}
    \label{table:hybrid score}
    \vspace{-15pt}
\end{table}

\section{Conclusion\label{sec: conclusion}}

This paper introduced a unified framework to derive safe control laws using energy functions. We proved that a variety of controllers can be derived from the unified framework by applying different hyperparameters. 
A benchmark system was introduced to evaluate the performance of different algorithms on a variety of scenarios with different system dynamics. The unified framework and the benchmark system helped us understand how hyperparameters of an algorithm affects the performance in terms of safety and efficiency. 
Based on the unified framework and comparison results, we proposed a new method, sublevel safe set algorithm (SSS). This new method combined the strengths of two best-performed algorithms: the safe set algorithm (SSA) and the barrier function algorithm (BFM), which outperform existing methods on most benchmark tests. 

\section*{Acknowledgement}
The authors would like to thank Jaskaran Grover, Hsien-Chung Lin, Liting Sun, and anonymous reviewers for their precious comments and suggestions.
\bibliography{main}

\begin{thebibliography}{10}
\providecommand{\url}[1]{#1}
\csname url@samestyle\endcsname
\providecommand{\newblock}{\relax}
\providecommand{\bibinfo}[2]{#2}
\providecommand{\BIBentrySTDinterwordspacing}{\spaceskip=0pt\relax}
\providecommand{\BIBentryALTinterwordstretchfactor}{4}
\providecommand{\BIBentryALTinterwordspacing}{\spaceskip=\fontdimen2\font plus
\BIBentryALTinterwordstretchfactor\fontdimen3\font minus
  \fontdimen4\font\relax}
\providecommand{\BIBforeignlanguage}[2]{{%
\expandafter\ifx\csname l@#1\endcsname\relax
\typeout{** WARNING: IEEEtran.bst: No hyphenation pattern has been}%
\typeout{** loaded for the language `#1'. Using the pattern for}%
\typeout{** the default language instead.}%
\else
\language=\csname l@#1\endcsname
\fi
#2}}
\providecommand{\BIBdecl}{\relax}
\BIBdecl

\bibitem{blanchini1999set}
F.~Blanchini, ``Set invariance in control,'' \emph{Automatica}, vol.~35,
  no.~11, pp. 1747--1767, 1999.

\bibitem{khatib1986real}
O.~Khatib, ``Real-time obstacle avoidance for manipulators and mobile robots,''
  in \emph{Autonomous robot vehicles}.\hskip 1em plus 0.5em minus 0.4em\relax
  Springer, 1986, pp. 396--404.

\bibitem{gracia2013reactive}
L.~Gracia, F.~Garelli, and A.~Sala, ``Reactive sliding-mode algorithm for
  collision avoidance in robotic systems,'' \emph{IEEE Transactions on Control
  Systems Technology}, vol.~21, no.~6, pp. 2391--2399, 2013.

\bibitem{ames2014control}
A.~D. Ames, J.~W. Grizzle, and P.~Tabuada, ``Control barrier function based
  quadratic programs with application to adaptive cruise control,'' in
  \emph{53rd IEEE Conference on Decision and Control}.\hskip 1em plus 0.5em
  minus 0.4em\relax IEEE, 2014, pp. 6271--6278.

\bibitem{liu2014control}
C.~Liu and M.~Tomizuka, ``Control in a safe set: Addressing safety in
  human-robot interactions,'' in \emph{ASME 2014 Dynamic Systems and Control
  Conference}.\hskip 1em plus 0.5em minus 0.4em\relax American Society of
  Mechanical Engineers, 2014, pp. V003T42A003--V003T42A003.

\bibitem{branicky1998multiple}
M.~S. Branicky, ``Multiple lyapunov functions and other analysis tools for
  switched and hybrid systems,'' \emph{IEEE Transactions on automatic control},
  vol.~43, no.~4, pp. 475--482, 1998.

\bibitem{berkenkamp2017safe}
F.~Berkenkamp, M.~Turchetta, A.~Schoellig, and A.~Krause, ``Safe model-based
  reinforcement learning with stability guarantees,'' in \emph{Advances in
  neural information processing systems}, 2017, pp. 908--918.

\bibitem{sun2018fast}
L.~Sun, C.~Peng, W.~Zhan, and M.~Tomizuka, ``A fast integrated planning and
  control framework for autonomous driving via imitation learning,'' in
  \emph{ASME 2018 Dynamic Systems and Control Conference}.\hskip 1em plus 0.5em
  minus 0.4em\relax American Society of Mechanical Engineers, 2018, pp.
  V003T37A012--V003T37A012.

\bibitem{maciejewski1985obstacle}
A.~A. Maciejewski and C.~A. Klein, ``Obstacle avoidance for kinematically
  redundant manipulators in dynamically varying environments,'' \emph{The
  international journal of robotics research}, vol.~4, no.~3, pp. 109--117,
  1985.

\bibitem{mitchell2005time}
I.~M. Mitchell, A.~M. Bayen, and C.~J. Tomlin, ``A time-dependent
  hamilton-jacobi formulation of reachable sets for continuous dynamic games,''
  \emph{IEEE Transactions on automatic control}, vol.~50, no.~7, pp. 947--957,
  2005.

\bibitem{macek2002reinforcement}
K.~Macek, I.~PetroviC, and N.~Peric, ``A reinforcement learning approach to
  obstacle avoidance of mobile robots,'' in \emph{7th International Workshop on
  Advanced Motion Control. Proceedings (Cat. No. 02TH8623)}.\hskip 1em plus
  0.5em minus 0.4em\relax IEEE, 2002, pp. 462--466.

\end{thebibliography}
\bibliographystyle{IEEEtran}

\begin{appendix}
\subsection{Proof of \cref{main theorem}}
\begin{proof}
    
    We prove the claim case by case.
    
    \begin{case}
        We first show that non-optimization based methods, PFM in \eqref{eq: pfc} and SMA in \eqref{eq: smc}, satisfy the claim.
        
        PFM provides control inputs in the Cartesian space, but what we need is the control input in the configuration space. In order to get the control input $\v{u}$, we derive the following equation based on \eqref{eq: c_h_x} and \eqref{eq: dynamics},
        \begin{align}
            \dtv{c}_r = \v{J}_{\v{c}_r}\ \dtv{x} = \v{J}_{\v{c}_r}\ (\v{f} + \v{g}\ \v{u}). \label{eq: PFM_3}
        \end{align}
        Combining \eqref{eq: PFM_1}, \eqref{eq: PFM_2}, and \eqref{eq: PFM_3}, we have
        \begin{align}
            \v{J}_{\v{c}_r}\ (\v{f} + \v{g}\ \v{u}) = \v{u}_c = \v{J}_{\v{c}_r}\ \v{g}\ \v{u}_0 - c_1\ \nabla \tilde \phi
        \end{align}
        
        In most cases, $\v{J}_{\v{c}_r}$ and $\v{g}$ are not invertible. Thus, we design a cost function $L$ to measure the difference between $\v{J}_{\v{c}_r}\ \v{g}\ \v{u}$ and $\v{u}_c$ and try to find $\v{u}$ by gradient descent:
        \begin{align}
            L &= \frac{1}{2} \norm{\v{J}_{\v{c}_r}\ (\v{f} + \v{g}\ \v{u}) - \v{J}_{\v{c}_r}\ \v{g}\ \v{u}_0 + c_1\ \nabla \tilde \phi}^2,\\
            \pdv{L}{u} &=  (\v{J}_{\v{c}_r}\ \v{g}) \transpose\ [\v{J}_{\v{c}_r}\ (\v{f} + \v{g}\ \v{u}) - \v{J}_{\v{c}_r}\ \v{g}\ \v{u}_0 + c_1\ \nabla \tilde \phi]. 
        \end{align}
        Initially, $\v{u} = \v{u}_0$. Hence
        \begin{align}
            \pdv{L}{u} &= (\v{J}_{\v{c}_r}\ \v{g}) \transpose\ [\v{J}_{\v{c}_r}\ \v{f} + c_1\ \nabla \tilde \phi]\\
            &= (\v{J}_{\v{c}_r}\ \v{g}) \transpose\ \v{J}_{\v{c}_r}\ \v{f} + c_1\ \v{g}\transpose\ \v{J}_{\v{c}_r}\transpose\ \nabla \tilde \phi\
        \end{align}
        Since we are using double integrator models, $(\v{J}_{\v{c}_r}\ \v{g}) \transpose\ \v{J}_{\v{c}_r}\ \v{f} = 0$. Based on \eqref{eq: phi_x_c}, we have
        \begin{align}
            \pdv{L}{u} &= c_1\ \v{g}\transpose\ \v{J}_{\v{c}_r}\transpose\ \nabla \tilde \phi = c_1\ \LgP\transpose.
        \end{align}  
        We use a unit step size for gradient and get
        \begin{align}
            \v{u} &= \v{u}_0 - \pdv{L}{u}= \v{u}_0 - c_1\ \LgP \transpose.
        \end{align}
        Therefore, \eqref{eq: pf_a} is verified and $\beta = 0$ for PFM.
        

        For SMA in \eqref{eq: smc}, 
        we can easily get
        \begin{align}
            \begin{cases}
                \v{u}^s = \v{u}_0^s - \mathbb{I}_B(\phi)\  c_2 \LgP\\
                \v{u}^e = \v{u}_0^e
            \end{cases}.
        \end{align}
        Therefore, \eqref{eq: sm_a} is verified and $\beta = 0$ for SMA.
        
    \end{case}
    
    \begin{case}
        We then show that optimization based methods, SSA in \eqref{eq: SSA} and BFM in \eqref{eq: BFM}, satisfy the claim. The constraints regarding $\dot\phi$ in these two methods are equivalent to \eqref{eq: constrain0}, except that the constraint in SSA is only effective when $\phi\geq 0$. In SSA in \eqref{eq: SSA}, $\xi = \eta$. In BFM in \eqref{eq: BFM}, $\xi = \lambda\phi$. Considering the perpendicular decomposition of the control input, the constraint \eqref{eq: constrain0} can be written as 
        \begin{equation}\label{eq: inequality of phi dot}
	\LgP\ \v{u}^s  \leq \xi - \LfP = \gamma \norm{\LgP}^2.
	\end{equation}
        The optimization objective $\norm{\v{u}_0 - \v{u}}^2$ is equivalent to $\norm{\v{u}_0^s - \v{u}^s}^2 + \norm{\v{u}_0^e - \v{u}^e}^2$. The minimization is achieved at
        \begin{align}
            \v{u}^e &= \v{u}_0^e,\\
            \v{u}^s &=\left\{\begin{array}{ll}
                \gamma\ \LgP\transpose& \mathrm{if } \mu  > \gamma \\
                \mu\ \LgP\transpose & \mathrm{otherwise}
            \end{array}\right..\label{eq: SSA and BFM parallel control}
        \end{align}
        Hence in both SSA and BFM, $\beta = 0$. 
        
        In SSA in \eqref{eq: SSA}, the safe control deviates from the reference control only when $\phi\geq 0$ and $\mu  > \gamma$, which corresponds to the case that $\mathbb{I}_A(\gamma)\ \mathbb{I}_B(\phi) = 1$. Rearranging \eqref{eq: SSA and BFM parallel control} into the form in \eqref{eq: safe control general form}, we can verify $\alpha$ in \eqref{eq: ss_a}.
        
        In BFM in \eqref{eq: BFM}, the safe control deviates from the reference control only when $\mu  > \gamma$, which corresponds to the case that $\mathbb{I}_A(\gamma) = 1$. Rearranging \eqref{eq: SSA and BFM parallel control} into the form in \eqref{eq: safe control general form}, we can verify $\alpha$ in \eqref{eq: bf_a}.
    \end{case}
    
\end{proof}

\subsection{Robot Dynamic Models}
BIS currently include four different robot dynamic models: ball robot model, unicycle robot model, SCARA robot model, and 4 DoF robot arm robot model. Their states and system dynamics are listed below.

\subsubsection{2D Ball}
The state of a 2D ball is composed of Cartesian position and velocity. Control inputs are accelerations.
\begin{align*}
    \v{x} = \begin{bmatrix}
        c_x \\
        c_y \\
        v_x \\
        v_y
    \end{bmatrix}
    \v{f} = \begin{bmatrix}
        v_x \\
        v_y \\
        0 \\
        0
    \end{bmatrix}
    \v{g} = \begin{bmatrix}
        0 & 0\\
        0 & 0\\
        1 & 0\\
        0 & 1\\
    \end{bmatrix}
    \v{u} = \begin{bmatrix}
        \dot v_x\\
        \dot v_y
    \end{bmatrix}
\end{align*}

\subsubsection{Unicycle}
The state of a unicycle is composed of Cartesian position, velocity, and heading angle. Control inputs are accelerations and angular speeds.
\begin{align*}
    \v{x} = \begin{bmatrix}
        c_x \\
        c_y \\
        v \\
        \theta
    \end{bmatrix}
    \v{f} = \begin{bmatrix}
        v\ \cos(\theta) \\
        v\ \sin(\theta)\\
        0 \\
        0
    \end{bmatrix}
    \v{g} = \begin{bmatrix}
        0 & 0\\
        0 & 0\\
        1 & 0\\
        0 & 1\\
    \end{bmatrix}
    \v{u} = \begin{bmatrix}
        \dot v\\
        \dot \theta
    \end{bmatrix}
\end{align*}

\subsubsection{SCARA and 4 DoF Robot Arm}
A SCARA is actually a 2 DoF robot arm. The state of a robot arm is composed of joints angle vector $\boldsymbol{\theta}$ and joints angular speed vector $\boldsymbol{\dot \theta}$. Control inputs are angular accelerations vector $\boldsymbol{\ddot \theta}$. We give a unified description here:
\begin{align*}
    \v{x} = \begin{bmatrix}
        \boldsymbol{\theta}\\
        \boldsymbol{\dot \theta}
    \end{bmatrix}
    \v{f} = \begin{bmatrix}
        \boldsymbol{\dot \theta}\\
        \v{0}
    \end{bmatrix}
    \v{g} = \begin{bmatrix}
        \v{0}\\
        \v{I}
    \end{bmatrix}
    \v{u} = \begin{bmatrix}
        \boldsymbol{\ddot \theta}
    \end{bmatrix}
\end{align*}

\subsection{Trajectory Profile}

\begin{figure}[h]
    \centering 
    \includegraphics [width=0.4\textwidth]{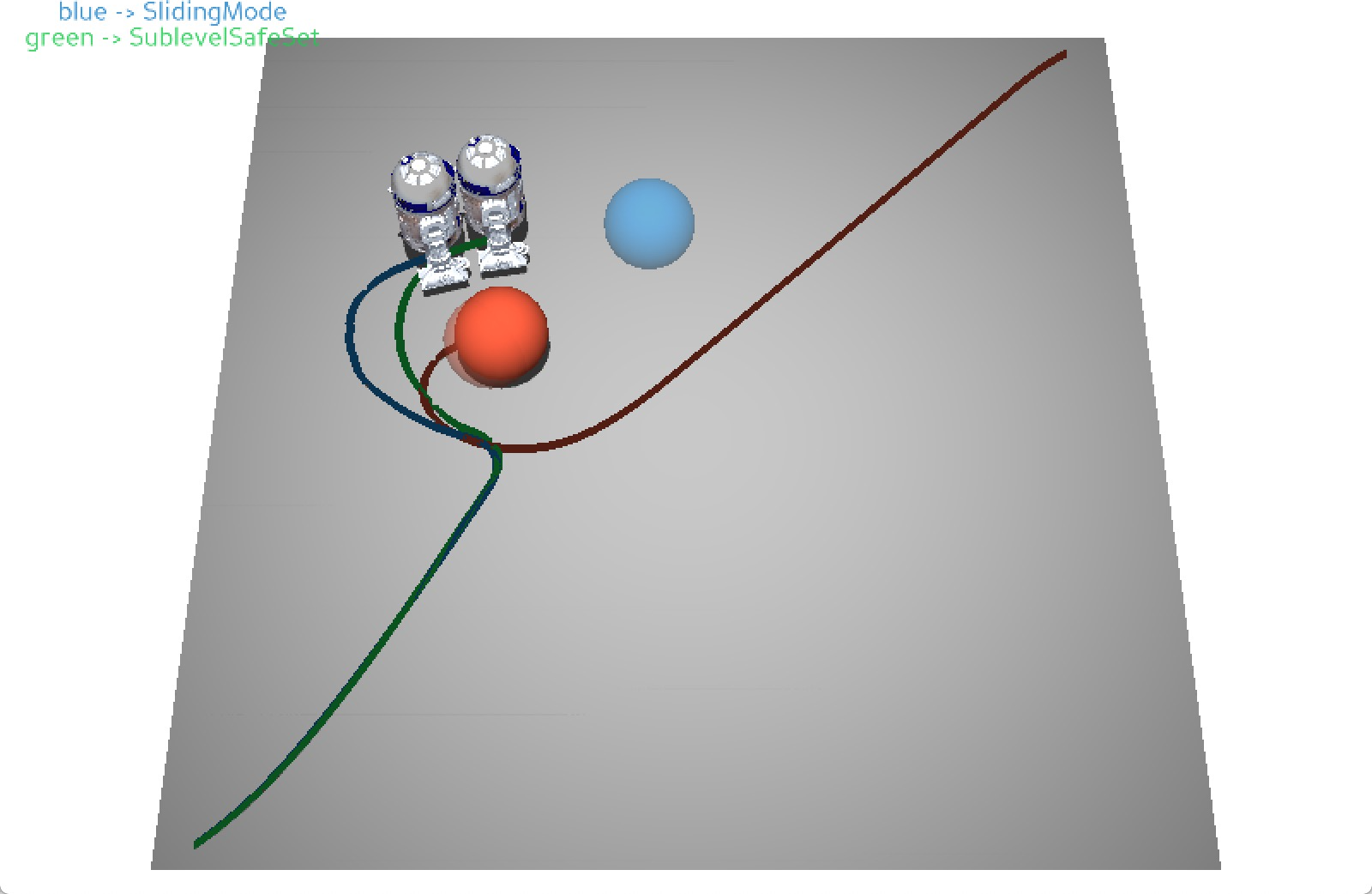}
    \caption{Simultaneous comparisons. The blue trajectory is generated by SMA and the green trajectory is generated by SSS.}
    \label{fig:visual comparison}
\end{figure}

To gain an intuition of the differences among algorithms, we visualize the trajectories of the robot under different control algorithms as shown in \cref{fig:visual comparison}. The environment setups and human trajectories are the same. We can see that SSS generates a more efficient trajectory than SMA.

\subsection{Discussion\label{sec: discussion}}
\subsubsection{Efficiency Compensation}
According to \cref{main theorem}, none of the methods discussed in this paper has an efficiency correction term, \ie, $\beta\equiv 0$. However, as we are using the simplest form of those methods, it does not imply that these methods cannot have an efficiency correction term. For example, if we change the objectives in SSA and BFM to be $(\v{u} - \v{u}_0)\transpose Q (\v{u} - \v{u}_0)$ for some positive definite $Q$ that is not identity, we can have a non trivial $\beta$. The detailed derivation and analysis of the $\beta$ term will be left for future work.  

It is worth noting that the efficiency correction term is useful for robots with kinematic redundancies, \eg, redundant robot arms. When the safety control correction is triggered, the robot's end effector may deviate from its desired trajectory to avoid collision between the robot link and the obstacle. For redundant robots, if the critical point is not at the end effector, the efficiency correction term can be adopted to make the end effector still follow its desired trajectory while allow the robot link to avoid collision. 
Maciejewski~\etal proposed an inverse-kinematics model in which the null space to the end effector is constrained to avoid link collisions \cite{maciejewski1985obstacle}. Our decomposition in \eqref{eq: u_decompose} is similar, though in an opposite way. Our approach finds an efficiency correction term in the null space of safety control that is associated with the critical point.

\subsubsection{Interactive and Non-Interactive Human Models}
In the benchmark, we used a non-interactive human model for the following two reasons.
\begin{enumerate}
    \item It is necessary to make sure all the controllers have exactly the same testing environment.
    \item Controllers should not count on the human subject for collision avoidance. It is necessary to validate the safety of the controllers with radical human subjects.
\end{enumerate}

\begin{figure}[t]
    \begin{center}
        \includegraphics[width=.4\textwidth]{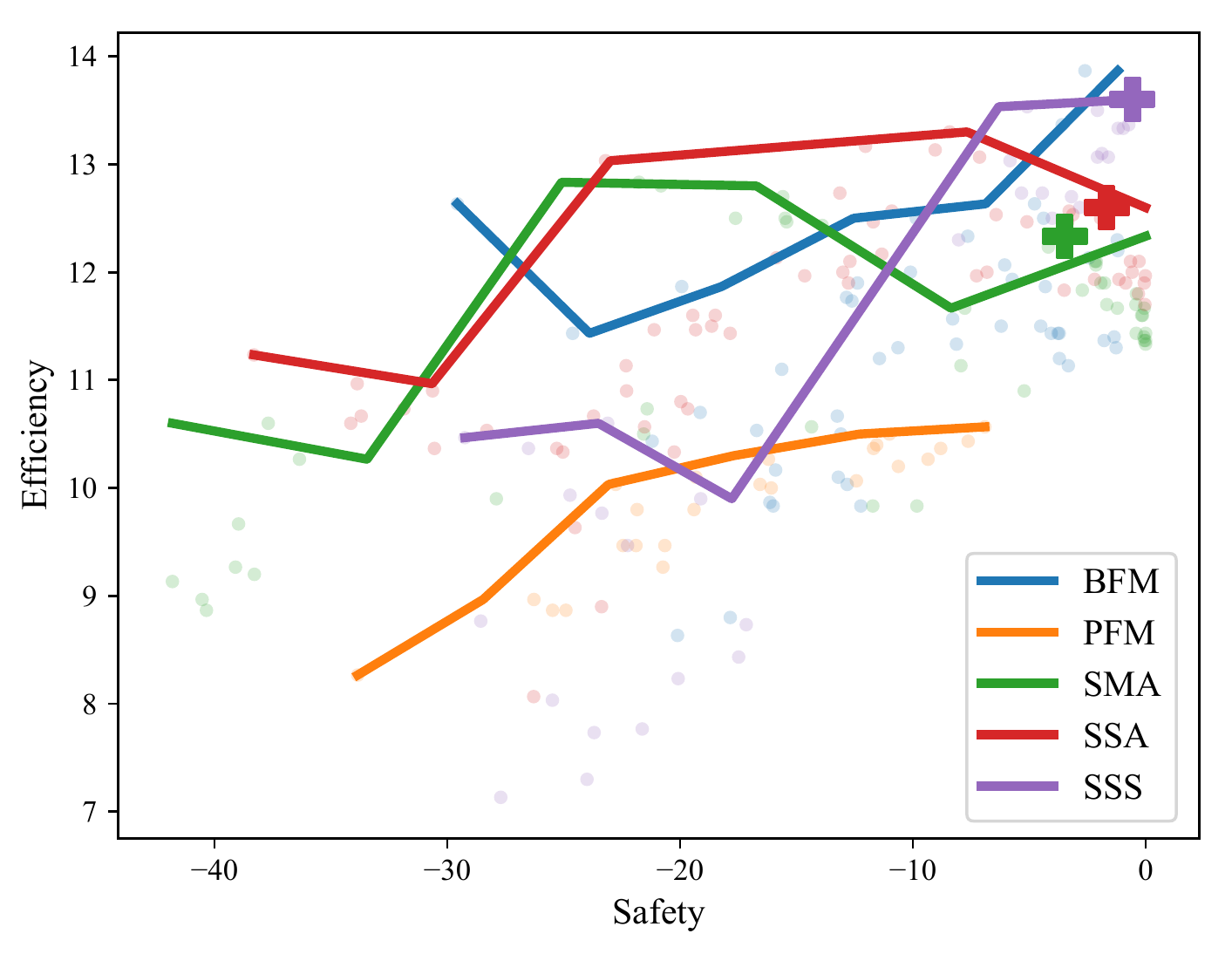}
        \caption{Trade off curves for interactive human model on SCARA platform}
        \label{fig:interactive result}
    \end{center}
\end{figure}

Nevertheless, we also conducted experiments with interactive human models to address a less restrictive scenario. In these experiments, since the behavior of human is affected by robot, we use the total goals achieved by both the human and the robot as the efficiency score. One of the results is shown in \cref{fig:interactive result}. The efficiency score is no longer negatively correlated to the safety score due to human's cooperation. In the future, we will rigorously analyze the performance of different safe control algorithms under interactive human behaviors.

\subsubsection{Related Methods}
Energy-function-based safe control methods only represent a portion of all safe control methods. Nonetheless, some safe control methods other than energy-function-based ones may implicitly incorporate functions that are similar to an energy function or a safety index. For example, the value function in the Hamilton-Jacobian backward reachability analysis~\cite{mitchell2005time}, and the value function in reinforcement learning methods~\cite{macek2002reinforcement}. These are numerical functions, while the safety indices in energy-function-based methods are analytical. Exploration of their connections are left for future works.

\end{appendix}

\end{document}